\documentclass[final,floatfix,twocolumn,pre,aps,floats,amsfonts,amssymb,showpacs]{revtex4}

\def\EQ{\begin{equation}}
\def\EN{\end{equation}}
\def\EQA{\begin{eqnarray}}
\def\ENA{\end{eqnarray}}

\begin{document}
\title{  
Scaling laws prediction from a solvable model\\
of turbulent thermal convection}
\author{ B. Dubrulle$^{1,2}$}

\address{
$^1$ NCAR, P.O. Box 3000, Boulder CO 80307-3000\\ 
$^2$ CNRS, URA 285, Observatoire
Midi-Pyr\'en\'ees, 14 avenue Belin, F-
31400 Toulouse, France
}

\begin{abstract}
A solvable turbulent model is used to predict both the structure
of the boundary layer
and the scaling laws in thermal convection. 
The transport of heat depends on 
the interplay between the thermal, viscous and integral scales of turbulence,
and thus, on both the Prandtl number and the Reynolds numbers. Depending
on their values, a wide variety of possible regimes is found, including
the classical $2/7$ and $1/3$ law, and a new 
$4/13=0.308$ law for the Nusselt power law variation with the 
Rayleigh number.
\end{abstract}


\date{Europhys. Letters, vol 51, 513-519 (2000)}

\maketitle

Until the advent 
of high precision numerical and 
experimental data, 
the heat transport in turbulent convection was thought
to be governed by the classical $1/3$ power law linking the Nusselt number
and the Rayleigh number. 
Deviations from this prediction have now been measured at
large Rayleigh numbers,
revealing a whole zoology of scaling exponents. They range from $0.2$ to $0.25$
in low Prandtl number, Mercury experiments \cite{CCS97}, to $0.29\pm 0.1$ 
in Helium
experiments \cite{Cast89,TSGS96,NSSD99}, while electro-chemical
convection  
gives the classical $1/3$ exponent \cite{Gols}. A number of 
theoretical models based on dimensional arguments
 have been developed to explain these new regimes
\cite{Cast89,ShraSigg90,CCS97}, including the possibility that no real scaling
prevails \cite{GrosLohs00}. 
In this letter, predictions obtained using a solvable
model of turbulent convection are presented. This model couples large scale
mean sheared velocity and temperature fields $U=(U(z),0,0)$ and 
$\Theta(z)$, with small scale random velocity and temperature fields.
This kind of large scale geometry is generally accepted as representative
of the Boussinesq convection in the boundary layers, due to the shearing effect 
of the large convective cells.
The model is closed by deriving an equation for the 
random component from the Boussinesq equation using two simplifying 
assumptions: i) the non-linear
interactions of the small scale scales 
between themselves is modeled via a turbulent viscosity; ii) the small
scale generation via the breaking of large scale structures 
is modeled by a random small scale forcing with prescribed statistics.
This results in a linear 
stochastic equation for the random small scales, which can be 
analytically solved in the shear flow geometry by a decomposition of
the small waves into localized wave-packets.  
This model was used to obtain analytical predictions of mean velocity
and/or temperature profiles in neutral boundary layer or
channel flows \cite{Naza99} or in the Planetary Surface Layer (PSL)
\cite{DLS99}.
Here, we adapt and generalize these results to determine both the 
structure of the boundary layers, and various scaling laws relating the
Nusselt number $Nu$ and various length scales
to the Rayleigh number $Ra$. Some of our findings are then 
compared with results from high resolution direct numerical 
simulations (DNS) at $Ra$ between $10^4$ and $10^8$, and at $Pr$ between $0.02$
and $7$, 
which are described in \cite{VerzCamu99,Kerr96,Wern93}.\

We consider Boussinesq equations, nondimensionalized by the thermal diffusivity
and the cell height:
\EQA
\partial_t u_i+\partial_j (u_i u_j)&=&-\partial_i p
+Ra Pr \theta \delta_{i3}+Pr\Delta u_i+f^{(u)}_i,\nonumber\\
\partial_t\theta+\partial_j (u_j\theta)&=&\Delta\theta+f^{(\theta)}. 
\label{constitutive}
\ENA
Here, $Pr$ is the Prandtl number, and $f^{(u)}$ and $f^{(\theta)}$ are small
scale random forces which are introduced to model the seeding
of small scales by turbulent plumes detaching from the wall and 
penetrating the outer region. For simplicity, this forcing is taken as 
spatially homogeneous and delta correlated in time
\cite{foot0}. These assumptions only influence the magnitude of the Reynolds
stresses, not their shape \cite{Naza99} . Averaging 
(\ref{constitutive}) over the statistics of the forcing and assuming
a shear flow geometry, one obtains
the standard equations  
for the x-component of ${\bf U}=<{\bf u}>=(U(z),0,0)$ and for 
$\Theta(z)=<\theta>$:
\EQA
\partial_t U+\partial_z<u'w'>
&=&-\partial_x P 
+Pr\partial_{z}^2 U,\nonumber\\
\partial_t \Theta
+\partial_z<w'\theta'> 
&=&
\partial_z^2\Theta.  
\label{LSEq}
\ENA
Here, the primes denote fluctuating (small scale) quantities and $<>$ the
averaging. In the stationary case, we get from (\ref{LSEq}) that $\partial_x P$
is a constant, independent of $z$. In the laminar case where 
$<u'w'>$ and $<w'\theta'>$ are negligible, we thus obtain the well known
parabolic profile for the velocity and the linear profile for the temperature.
In the turbulent case, the profiles are linear within the thermal or the
viscous layer, while outside this layer, they are given by the condition 
\EQ
\partial_z <u' w'>=-\partial_x P, \quad \partial_z <w'\theta'>=0.
\label{turbulentcase}
\EN   
To close the system, we need $<u'w'>$ and $<w'\theta '>$.
For this, we now derive an equation for the 
fluctuating quantities, by taking into account the scale separation between
the mean field and the random field $l/L=\eta\ll 1$. For this, we decompose
the velocity field into localized wave-packets via a Gabor transform (GT): 
\EQ
{\hat u}({\bf x},{\bf k},t)=\int g(\eta^\ast \vert
{\bf x-x'}\vert)e^{i{\bf k\cdot (x-x')}}
{\bf u}({\bf x'},t)d{\bf x'},
\label{gabordef}
\EN
where $g$ is a function which decreases rapidly at infinity and 
$1\gg \eta^\ast \gg\eta$. Because of this, the GT of any large 
scale field is exponentially small in $\eta^\ast$, and the GT
of $U_j\partial_j {\bf u'}_i$ can be developed into $U_j\partial_j \hat 
{u'}_i-\partial_j(U_m k_m)\partial_{k_j}\hat {u'}_i$ (see \cite{Naza99} for
details). Using these properties, we obtain after GT of (\ref{constitutive})
the small scale equation:
\EQA
D_t \hat u_i
&=&-ik_i \hat p- \hat w \partial_z U \delta_{i1}
+Ra Pr \hat \theta\delta_{i3}
-Pr^t {\bf k}^2 \hat u_i+\hat f^{(u)}_i\nonumber\\
D_t \hat \theta
&=&-\hat w\partial_z \Theta 
-{\bf k}^2\hat\theta+\hat f^{(\theta)}.
\label{ssEqmod}
\ENA
We have dropped primes on fluctuating quantities for convenient notations
and introduced the total derivative $D_t=
\partial_t +U\partial_x
-\partial_z (U k_x) \partial_{k_z}$ .
In (\ref{ssEqmod}), we have furthermore model the GT of
the non-linear terms describing local interactions between fluctuations
via the introduction of turbulent 
viscosity and diffusivity, or equivalently, via a turbulent 
Prandtl number $Pr^t$. 
The linear part of (\ref{ssEqmod}) is exact and describes
 the non-local interactions
between the mean and the fluctuating part. 
We have thus obtained a linear, stochastic equation for the fluctuating
quantities to close our system.\

To solve the closed system (\ref{LSEq}) and (\ref{ssEqmod}), we introduce 
two further simplification: i) we restrict ourselves to the 2D case. This
is justified because numerical work on 2D thermal convection showed that 
2D geometry is sufficient to capture the physical mechanism responsible 
for the $Nu$ scaling with $Ra$ \cite{Wern93}. Analytical
work on neutral shear flow has 
also proved that 2D geometry is sufficient to capture
the correct shape of the equilibrium profile \cite{Naza99}. Note that
the vortex stretching, which is theoretically absent in 2D 
geometry, has been implicitly accounted for via the turbulent
viscosity. In real 2D flows, this turbulent viscosity can be ignored
\cite{Naza99}.
ii) we take
$Pr^t=1$. This is because our model introduces a turbulent Prandtl number
whose exact value is unknown, which implies 
that the correct $Pr$ scaling cannot be captured
within our model. General $Pr$ effects will be introduced in another manner,
via links between length scales.
With these approximations, it was shown in \cite{DLS99}
that to leading order in the Reynolds number $Re$, 
the solutions of (\ref{ssEqmod})
satisfy:
\EQA
<w'u'>&=&
\frac{1}{2\partial_z U(a_+-a_-)^2}
\left(\frac{\partial_z U}{k_\ast^2}\right)
^{-(2+4a_+)/3} \nonumber \\
&& \Bigg(4a_-^2\lambda_1+\frac{Ra Pr}{\partial_z U}\lambda_2
+\left(\frac{Ra Pr}{\partial_z U}\right)^2\lambda_3\Bigg),\nonumber\\
<w'\theta'>&=&
\frac{1}{2\partial_z U(a_+-a_-)^2}
\left(\frac{\partial_z U}{k_\ast^2}\right)
^{-(2+4a_+)/3} \nonumber \\
&& \Bigg(-8a_+a_-^2\frac{\partial_z U}{Ra Pr}\mu_1+
4a_+a_-\mu_2
-2a_+\frac{RaPr}{\partial_z U}\mu_3\Bigg).
\label{expansionuwwt}
\ENA
Here, $k_\ast$ is a characteristic horizontal wavenumber, $\lambda_i$
and $\mu_i$ ($i=1,..3$) are some constant which depend on the forcing 
correlation functions, and $a_\pm=-(1\pm\sqrt{1-4 R_i})/2$, where 
\EQ
R_i=Ra Pr\frac{\partial_z \Theta}{\left(\partial_z U\right)^2},
\label{ridchardson}
\EN
is the Richardson number. In developed convective turbulence, this number is
large and negative $R_i\ll -1$. So only the leading order
contribution in $R_i$ in the expression (\ref{expansionuwwt}) will
be further considered. 
Combining (\ref{turbulentcase}) with the large $R_i$ expansion of 
(\ref{expansionuwwt}), we find that the turbulent profiles are given by
the conditions:
\EQA
<w' u'>&\approx& \frac{u_\ast^2}{\partial_z U}\propto z,\nonumber\\
<w'\theta'>&\approx& \frac{u_\ast^2 \sqrt{-R_i}}{RaPr}\propto cte=Nu,
\label{oufi}
\ENA
where we have introduced $u_\ast^2=(\partial_z U/k_\ast^2)^{2\sqrt{-R_i}/3}$.
Looking for simple solutions where $\partial_z U$ and $\partial_z \Theta$
are power laws of $z$, the only solution is:
\EQ
\partial_z U\sim \frac{1}{z},\quad \partial_z \Theta \sim \frac{1}{z^2},
\label{solutionprof}
\EN
resulting in a constant (with $z$) Richardson number. This solution corresponds
to the standard logarithmic velocity profile, and a temperature profile
decreasing like $1/z$. Such a profile had been predicted by Malkus \cite{Malk54}
 using a maximum principle. It was found to be in good agreement
with experiments of large $Ra$ convection \cite{Town59} and 
DNS at 
$Ra=10^8$ and $Pr=0.7$ \cite{Kerr96,KerrPC}. 
For the velocity profile,
the prediction (\ref{solutionprof}) is difficult to check numerically,
because the Reynolds number is too low for the boundary layer to be fully 
developed \cite{KerrPC}. Even in the large Raleigh number experiments
of \cite{Libchgroup}, the extent of the turbulent boundary layer 
is too small (a tenth of the cell) to check this prediction. However,
in the PSL, the prediction
(\ref{solutionprof}) appears compatible with the measurements \cite{DLS99}.\ 

It is also possible to find the leading order behavior in $Re$ and $R_i$
of vertical velocities and temperature fluctuations, using the results of 
\cite{DLS99}. They are:
\EQ
<w'^2>\approx R_i \frac{u_\ast^2}{\partial_z U},\quad
<\theta'^2>\approx R_i \frac{u_\ast^2\partial_z U}{(RaPr)^2}.
\label{tempetheta}
\EN
This give a r.m.s. vertical velocity and temperature varying like
$z^{1/2}$ and $z^{-1/2}$, like in the free-fall regime. For the temperature 
fluctuation, the predicted decrease is more rapid than the classical 
$z^{-1/3}$ prediction. This feature has been already observed in high 
$Ra$ convection \cite{Libchgroup}. 
In the PSL, the r.m.s. velocity and temperature
are compatible with (\ref{tempetheta}) \cite{DLS99}.\

From (\ref{oufi}) and (\ref{solutionprof}),
we may also derive interesting exact relations. By matching the 
turbulent profiles with the viscous or diffusive solutions
$Pr\partial_z U=u_\tau^2$ and $\partial_z \Theta=Nu$, where $u_\tau$ is the 
friction velocity and $Nu$ the Nusselt number, we get:
\EQ
\partial_z U=u_\tau^2 \frac{\lambda_V}{zPr},\quad \partial_z \Theta =Nu \left
(\frac{\lambda_T}{z}\right)^2,
\label{match1}
\EN
Here, we have introduced the thermal length scale $\lambda_T=1/Nu$ and 
the viscous length scale $\lambda_V=Pr/u_\tau$. 
To find the prefactor in (\ref{oufi}), we use the law of energy dissipation
in a boundary layer geometry which gives $\epsilon\sim <u'w'>\partial_z U$.
We then take into account the exact relation $\epsilon=Ra Nu$ and 
use ({\ref{oufi}), to obtain
$u_\ast^2=Ra Nu$ and:
\EQ
\frac{<u'w'>}{<w'\theta'>}=\frac{RaPr}{\partial_z U \sqrt{-R_i}}=
\frac{Ra}{\partial_z U}.
\label{utile2}
\EN
The first equality comes from (\ref{oufi}), while the second comes 
from the link between $\epsilon$ and $<u'w>$.
Condition (\ref{utile2}) shows that $R_i$ is independent of $Ra$. 
This condition is actually 
a necessary condition for stability of the turbulent boundary layer 
\cite{Cast89}. 
The definition (\ref{ridchardson}) combined with (\ref{match1}) then implies: 
\EQ
Ra=u_\tau^2 NuPr.
\label{exactutile}
\EN
We may also use (\ref{match1}) and the above estimate to define 
a characteristic vertical velocity $w_c=\sqrt{<w'^2>/z}$ and temperature
fluctuation $\Delta_c=\sqrt{<\theta'^2>z}$, which scale as:
\EQ
w_c\sim \sqrt{\frac{Ra Nu}{u_\tau}},\quad 
\Delta_c \sim \sqrt{\frac{u_\tau Nu }{Ra}}.
\label{defiwcandDc}
\EN
These exact relations derived within our nonlocal model will be the basis
of the scaling theory which we now detail.\

For this, we follow \cite{Kerr96} and introduce a third length scale,
noted $\lambda_I$, representing the location of the peak of the kinetic 
horizontal energy spectrum. Its expression can again be found within the 
nonlocal framework by noting that this length scale coincides with the location 
of the maximum of the r.m.s. horizontal velocity. In the non-local
model, this horizontal velocity is passively advected by the large scale
horizontal velocity and obeys $U\partial_x u'=\partial_z^2 u'$. The solution
to this equation depends on the structure of the velocity boundary layer:
if the Reynolds number is too small, the boundary layer is mainly laminar,
$U=u_\tau z/\lambda_V$ and simple dimensional argument show that
$u'$ is a function of $z/(x\lambda_V u_\tau^{-1})^{1/3}$. For a fixed
aspect ratio $\Gamma$, this defines a 
typical vertical scale of r.m.s. velocity variation $\lambda_I=(\Gamma\lambda_V 
u_\tau^{-1})^{1/3}$. 
When the Reynolds number exceeds
a critical value of the order of $Re_c\sim 10^5$ \cite{Schl}, the
boundary layer turbulence is well developed, and most of the transport of
$u'$ is provided by the turbulent logarithmic 
regime $U\sim u_\tau (ln z +B)$. This regime is hard to obtain
at Prandtl numbers of the order one or larger 
and it is likely that only low Prandtl
number system (like Mercury experiments) have the 
ability to reach this critical Reynolds number \cite{CCS97}.
In this regime, $u'$ is a function 
of $z/(x u_\tau^{-1})^{1/2}$, thereby defining a typical length scale 
$\lambda_I\sim (\Gamma u_\tau^{-1})^{1/2}$ \cite{foot2}. The two regimes can be
lumped into the single formula $\lambda_I\sim \lambda_V (\Gamma u_\tau
Pr^{-2})^\psi$, where $\psi=1/3$ for $Re<Re_c$ and $\psi=1/2$ for
$Re>Re_c$.
Using these results, the logic of our argumentation
is now to link $u_\tau$ to $Nu$ by relating $\lambda_V$ and/or 
$\lambda_I$ to $\lambda_T$, and then use the exact relation (\ref{exactutile})
to obtain $Nu$ versus $Ra$. For this, we set 
$\lambda_T=\lambda_V^\alpha \lambda_I^{1-\alpha}$, where $0\le \alpha\le 1$ is
a parameter which enables to single out three remarkable regimes. 
In the first one, $\alpha=1$, $\lambda_T=\lambda_V$: the viscous and thermal
layer coincide. This situation might be typical of large aspect ratio cells,
or in numerical simulations with stress-free boundary conditions
\cite{WernPC}. In our model, it corresponds to $Nu\sim Ra^{1/3} Pr^{-1}$,
the classical case. 
In the second regime, $0<\alpha<1$ and the length-scale ordering changes to
$\lambda_V<\lambda_T<\lambda_I$. This situation seems to be typical 
for convection at $Pr\sim 1$ \cite{Kerr96}. For illustration 
purpose, it is interesting to single out two special value of $\alpha$
which are relevant in turbulence:  
the first one is 
$\alpha=2/3$. This corresponds to $\lambda_T$ being the Taylor micro-scale,
given by the square root of the mean energy to the mean enstrophy. The second 
value is $\alpha=2/5$. The corresponding scale varies like 
$\lambda_I (\lambda_I Re)^{-3/10}$, where $Re$ is the Reynolds number based 
on the cell size and on the  velocity at the integral scale $U_I$.
It was shown \cite{MTW99} to correspond to the maximum of the Kolmogorov
function and represents the location of the middle of the inertial range.
At last, the third regime is obtained with $\alpha=0$, making 
$\lambda_T=\lambda_I$. This situation is typical of low Prandtl number 
convection. 
In our estimate of $\alpha$, we have followed standard
turbulence phenomenology, and assumed that 
$\lambda_I=\lambda_V (\lambda_I Re)^{3/4}$. 
 Note that for
the heat transport, the exact relation in the two regimes
($\psi=1/3$ or $1/2$) is:
\EQ
Nu\sim Ra^{\frac{1-\psi[1-\alpha]}{3-\psi[1-\alpha]}}
\Gamma^{-\frac{2\psi[1-\alpha]}{3-\psi[1-\alpha]}} 
Pr^{\frac{5\psi[1-\alpha]-3}{3-\psi[1-\alpha]}}.
\label{tout}
\EN
The Prandtl dependence obtained in our model is stronger than
in standard models of turbulent convection. It is not necessarily 
inconsistent with available data.
Such a dependence would for example account for most of the difference between
the $Pr=4$ experiment of \cite{ahlers} and the $Pr=0.8$ experiment
of \cite{AshkStein99}.
However, we do not expect our model to fully capture the general
Prandtl dependence, because of our approximation 
on the 
turbulent Prandtl number. In the sequel, we thus concentrate on the Rayleigh
dependence of the physical quantities.
The various scaling exponents predicted by the combination of (\ref{exactutile})
, (\ref{defiwcandDc})
and the length scale relation is summarized in Table I. 
For purpose
of comparison with the DNS,
we have also included the predictions for $Re<Re_c$ 
at three special values of 
$\alpha$, relevant to the low and order unity Prandtl number.\
\begin{table}
\begin{tabular}{c|c|c|c|c}\hline
\multicolumn{1}{c|}{Name} &\multicolumn{1}{c|}{General case}
&\multicolumn{3}{c|}{Value for $Re<Re_c$ ($ \psi=1/3$)}\\
\multicolumn{1}{c|}{}
&\multicolumn{1}{c|}{} &\multicolumn{1}{c|}{$\alpha=0$}
&\multicolumn{1}{c|}{$\alpha=2/3$}
&\multicolumn{1}{c}{$\alpha=2/5$}\\ \hline
$Nu$ 
&$\frac{1-\psi(1-\alpha)}{3-\psi(1-\alpha)}$ 
&$0.2500$
&$0.3077$ &$0.2857$\\
$Lw_c/\nu$ 
&$\frac{1}{2}\frac{3-2\psi(1-\alpha)}{3-\psi(1-\alpha)}$ 
&$0.4375$ &$0.4808$ &$0.4643$\\
$\Delta_c$ 
&$-\frac{1}{2}\frac{1}{3-\psi(1-\alpha)}$ 
&$-0.1875$ &$-0.1731$ &$-0.1786$\\
$u_\tau$ 
&$\frac{1}{3-\psi(1-\alpha)}$ 
&$0.3750$
&$0.3462$ &$0.3571$\\
$\lambda_V$ 
&$-\frac{1}{3-\psi(1-\alpha)}$ 
&$-0.3750$
&$-0.3462$ &$0.3571$\\
$\lambda_T$ 
&$-\frac{1-\psi(1-\alpha)}{3-\psi(1-\alpha)}$ 
&$-0.2500$
&$-0.3077$ &$-0.2857$\\
$\lambda_I$ 
&$\frac{\psi-1)}{3-\psi(1-\alpha)}$ 
&$-0.2500$
&$-0.2308$ &$-0.2381$\\ \hline
\end{tabular}
\label{table1}
\caption{Summary of exponents as a function of the
Rayleigh number as predicted by the present model. Here,
$\psi=1/3$ when $Re<Re_c$ and $\psi=1/2$ when $Re>Re_c$.
$0\le\alpha\le 1$ is a parameter describing how
the thermal
length-scale compares with the viscous $\lambda_V$ or integral scale
$\lambda_I$, thereby describing different Prandtl regimes, like
$\alpha=0$ for low Prandtl, and $\alpha=2/3$ or $2/5$ for $Pr\sim 1$.
$w_c$ and $\Delta_c$ are characteristic 
r.m.s. vertical velocity and temperature.}
\end{table}

In the low Prandtl regime, our model predicts a transition between
a $Nu=Ra^{1/4}$ regime up to $Nu=Ra^{1/5}$ regime at larger $Re$ or $Ra$.
This is in agreement with the experimental findings of \cite{CCS97}.
In the regime of $Pr\sim 1$, the prediction depends on the value of $\alpha$.
At $Re<Re_c$,
we find $\beta=4/13$
($\alpha=2/3$) and $\beta=2/7$ ($\alpha=2/5$).
The two values are close to the $0.29$ value usually observed in experiments
or in simulations. The $2/7$ value is here exactly recovered when
the thermal length scale matches the middle of the inertial range, which
may help give a novel interpretation of the $2/7$ law. The 
value $\beta=4/13=0.3077$ is in remarkable agreement with the experimental
value $\beta=0.309\pm 0.0043$ obtained in 
\cite{NSSD99} using an Helium experiment with $0.5$ aspect ratio
and spanning eleven decades of 
Rayleigh number. The case where 
$Re>Re_c$ gives slightly different values.
They are respectively $\beta=5/17=0.294$ ($\alpha=2/3$) and $\beta=7/27=0.259$
($\alpha=2/5$). The first value is so close to the value obtained before
the transition, that it would make any transition difficult 
to detect experimentally. This may explain why no transition was detected in
\cite{NSSD99}, 
despite the large Rayleigh values attained. The
second value is sufficiently lower than the value before transition, so
that it should be detectable in a careful experiment.
 Finally,
in the third regime, we find $\beta=1/3$ at all Reynolds number, in agreement 
with the $Pr=1$ DNS of \cite{Wern93}. Our model
seems to rule out the $\beta=1/2$ which has been 
predicted by Kraichnan \cite{Krai62} 
at very large Rayleigh number, and which might have been experimentally 
detected by Chavanne {\sl et al}
\cite{CCCHCC97}. If this regime is confirmed, it might then correspond
to a situation where the  boundary layer grows unstable, thereby unvalidating
a starting assumption of our model. 
For $w_c$ and $\Delta_c$  scaling with $Ra$,
very good agreement between the prediction and the experimental 
value $0.43$ and $-0.2$ \cite{CCS97}
 is obtained at low Prandtl number. For $Pr\sim 1$, our estimates
of $w_c L/\nu$ coincides 
for $\alpha=2/3$ with the value $0.485\pm 0.005$ 
measured in Helium \cite{Cast89}, while the exponent for $\Delta_c$
is slightly larger than the experimental value $-0.147$.\ 

Another test  of our predictions can also be made by comparison
with DNS, where some length scales have been 
measured. The results are shown in Table II. They are in
rather good agreement with the theoretical result, except for the case 
$Pr=7$ of \cite{Kerr96} where both the non-dimensional vertical
velocity and the integral scale exponent deviate substantially from 
the prediction. It would certainly be interesting to investigate further 
the meaning of this discrepancy. For instance, it might be due to 
a Rayleigh dependence of the turbulent Prandtl number. 
Finally, we note that the aspect ratio is
also likely to modify the recirculation pattern within a given cell 
and change the relative behavior between the length scales, so different
scaling may appear at different aspect ratio, as observed in \cite{ahlers}.\

I thank Robert Kerr and Joseph Werne for references and interesting discussions.
I have been
partially funded by a NATO fellowship.

\begin{table}
\begin{tabular}{c|c|c|c|c|c}\hline
\multicolumn{1}{c|}{Name} &\multicolumn{2}{c|}{Verzicco and Camussi}
&\multicolumn{1}{c|}{Werne (2D)}
&\multicolumn{2}{c}{Kerr and Herring}\\
\multicolumn{1}{c|}{}&\multicolumn{1}{c|}{$Pr=0.02$}
&\multicolumn{1}{c|}{$Pr=0.7$} &\multicolumn{1}{c|}{$Pr=7$}
&\multicolumn{1}{c|}{$Pr=0.07$}
&\multicolumn{1}{c}{$Pr=7$}\\ \hline
$Lw_c/\nu$ &$$ &$$ &$0.54$
&$0.47$ &$0.56$\\
$u_\tau$ &$$ &$$ &$0.39$
&$$ &$$\\
$\lambda_T$ &$-0.25$ &$-0.29$ &$$
&$-0.26$ &$-0.29$\\
$\lambda_I$ &$-0.18$ &$-0.23$ &$$
&$-0.26$ &$-0.11$\\\hline
\end{tabular}
\label{table2}
\caption{Summary of exponents as a function of the
Rayleigh number as measured in various DNS at aspect ratio
 1 for the three first cases, and 4 for the last two.}
\end{table}


\begin{thebibliography}{99}
\bibitem{foot0}
Similar results would be obtained using a shortly correlated forcing.
\bibitem{foot2}
A similar line of argumentation 
was done by \cite{CCS97} for $\lambda_T$, using the transport of 
$\Theta$ instead of $u'$.
\end{thebibliography}
\end{document}